\documentclass{ws-procs9x6}

\begin{document}

\title{Non-standard baryon-dark matter interactions}

\author{B. FAMAEY$^{1,2*}$, J.-P. BRUNETON$^3$}

\address{$^1$IAA, Universit\'e
Libre de Bruxelles, 1050 Bruxelles, Belgium \\
$^2$AIfA, Universit\"at Bonn, 53121 Bonn, Germany\\
$^3$SISSA/ISAS, Via Beirut 2--4, 34014, Trieste and INFN Sezione di Trieste, Via Valerio, 2, 34127 Trieste, Italy\\
$^*$E-mail: bfamaey@astro.uni-bonn.de}

\begin{abstract}
After summarizing the respective merits of the Cold Dark Matter (CDM) and Modified Newtonian Dynamics (MOND) paradigms in various stellar systems, we investigate the possibility that a non-standard interaction between baryonic and dark matter could reproduce the successes of CDM at extragalactic scales while making baryonic matter effectively obey the MOND field equation in spiral galaxies.
\end{abstract}


\bodymatter

\section{Introduction}

Data on large scale structures point towards a Universe dominated by dark matter and dark energy. Discovering the nature of these mysterious components of the Universe is, without a doubt, the major challenge of modern astrophysics. Nowadays, the dominant paradigm is that dark matter is made of non-baryonic weakly interacting massive particles, the so-called {\it cold dark matter} (CDM), and that dark energy is well represented by a cosmological constant ($\Lambda$) in Einstein equations. The $\Lambda$CDM cosmological model has known a remarkable success in explaining and predicting diverse data sets corresponding to the Universe at its largest scales. Nevertheless, an ever growing number of observations at galactic scales are in conflict with the predictions of the $\Lambda$CDM cosmological model. One often assumes that these problems are due to the fact that baryons are playing an important role in the dynamics of galaxies, and that their physics is not incorporated rigorously enough in cosmological simulations down to galactic scales\cite{Mashchenko}. However, a standard feedback from the baryons in the form of supernova explosions and stellar winds would require an extreme fine-tuning to explain the conspiracy between the baryonic and DM distributions in spiral galaxies, which is actually {\it independent} of the particular history of each stellar system. Signatures of this conspiracy are (i) the baryonic Tully-Fisher (BTF) relation\cite{McGaugh}, (ii) the fact that the mass discrepancy in galaxies always appears at the same gravitational acceleration $a_0 \sim 10^{-10}{\rm m}/{\rm s}^{2}$ (an acceleration also defining the zero-point of the forementioned BTF relation),  or (iii) the fact that galaxy rotation curves display obvious features (bumps or wiggles) that are also clearly visible in the stellar or gas distribution. In short, a one-to-one correspondence appears to exist between the integrated baryonic mass and the integrated dark mass at all radii in rotating galaxies, and not only in the outskirts. This relation is known as {\it Milgrom's law}\cite{Milgrom,SMc}, and can be interpreted as a modification of Newtonian dynamics (MOND). 

Although MOND is very successful at galactic scales, this phenomenological law encounters some problems at larger scales. It is now well established that interpreting the data using MOND's law does not completely remove the need for DM in clusters of galaxies. A large amount of unseen matter is still necessary to fully explain both lensing and dynamics of these objects. Another problem related to MOND's phenomenology is that it still lacks a well-defined relativistic formulation\cite{JPBGEF}, that would be both natural (free of fine-tunings) and consistent from a field theory point of view. Related to this is the fact that the physical origin of a new acceleration scale, and of the modification of the physical laws that it implies, are rather unclear at present. Some very interesting numerical coincidences ($a_0 \sim H_0 \sim \Lambda^{1/2}$) may pave the way to some unified picture, but to date, no major workable ideas have been proposed in this direction. 

Having therefore noticed (i) the difficulty to encapsulate the MOND phenomenology in a fully consistent theory, (ii) the abscence of clear physical origin to it, and (iii) the fact that the phenomenology anyway fails at extragalactic scales, we are naturally led to the idea that MOND could be only an effective picture, valid at galactic scales, of a more general theory. We will therefore explore the possibility that the modification of gravity assumed in MOND may originate from a new interaction between DM and baryons, that could explain Milgrom's law at galactic scales while also reproducing the successes of CDM at larger scales.

Section 2 deals more precisely with the successes and problems of the CDM and MOND paradigms at different scales. This will help us to discuss in Section 3 some model-independent considerations about the possibilty of a unification of CDM and MOND. We will notably show that some new dimensionful constants (besides $a_0$) may be needed to unify properly these two paradigms in a single framework. We will finally present a class of models that encapsulate most of the desired phenomenology, although this should be viewed only as a very first step, as the resulting model does suffer from naturalness problems, and is still far from explaining the physical origin of $a_0$.

\section{CDM vs. MOND}

In this Section, we compare the respective merits of CDM and MOND in various stellar systems and summarize this in a table, $+$ and $-$ signs being attributed to each paradigm, with the inherent subjectivity that this kind of classification encompasses. Obviously, one could argue that the concordance $\Lambda$CDM model is a fully defined cosmological model, and that putting it on equal footings with a simple phenomenological law is unfair. On the other hand this phenomenological law is fully predictive in stationary systems, and its lack of present-day physical basis can be considered as less harmful than the fine-tuning problems that the CDM paradigm encounters in galaxies. Let us also note that the actual situation is often much more complex than what can be summarized with a straightforward $+$ or $-$ sign. We take here the most pessimistic attitude, which is to put a $-$ sign whenever a paradigm encounters a single problem: this results in a globally very pessimistic overview of both paradigms, but it does not necessarily mean that these problems cannot be solved within each paradigm. With all these important caveats in mind, we proceed hereafter to the comparison.

\subsection{Galactic scales}

In high surface brightness (HSB) spiral galaxies (such as the Milky Way\cite{MW,Bienayme}) MOND predictions are extremely successful\cite{SMc} while the CDM paradigm suffers from obvious problems such as the cusp and the fine-tuning problem resulting from MOND's success. What is more, the transfer of angular momentum from the disk to the dark halo prevents from forming large enough baryonic disks in the process of galaxy formation (the angular momentum problem), a problem which is not present in MOND. In low surface brightness spiral galaxies (LSB), the situation is even worse for CDM, where the bumps and wiggles of the rotation curves following the light distribution are hard to understand. Let us note that the high mass-discrepancy in these systems was predicted by MOND before the first LSB rotation curves were measured. What is more, MOND predictions are also extremely successful in rotationally supported dwarf irregular (dIrr) galaxies.

For tidal dwarf galaxies (TDG), formed in the tidal tails resulting from the collision of two spiral galaxies, CDM simulations predict that the objects should be devoid of collisionless DM since they are formed out of the material that was in the rotating disks of the progenitors. However, rotation curves of three TDGs in the NGC~5291 system indicate the presence of large amounts of DM\cite{Bournaud} (if one agrees that the HI envelopes are bound to the TDGs). The solution for CDM would then be to assume large amounts of baryonic DM in the disks of the progenitors. However the mass-discrepancy in these TDGs perfectly obeys the MOND law\cite{Gentile}, which would thus require a really {\it ad hoc} distribution of baryonic DM, similar to the fine-tuned distribution of CDM in other rotationally supported galaxies. 

On the other hand, while the velocity dispersion profiles of dwarf spheroidal (dSph) satellites of the Milky Way are generally in accordance with MOND, Sextans and Draco, as well as the new ultra-faint dSphs, would require unrealistically high stellar mass-to-light ratios\cite{Angus1}, thus meaning that these galaxies may be problematic for MOND (if one agrees that all stars used for computing their velocity dispersion profiles are bound and in equilibrium). Let us however note that they are also problematic for the CDM paradigm, since they also suffer from the cusp problem, and their planar distribution around the Milky Way could argue as them actually being old TDG relics rather than CDM building blocks\cite{Kroupa}. 

Finally, in some large elliptical galaxies, the dearth of DM in the central parts can be problematic for the CDM paradigm\cite{Tiret}, but on the other hand some ellipticals (essentially at the center of clusters) require unrealistically high mass-to-light ratio in the MOND context\cite{Richtler,Shan}. 

\subsection{Subgalactic scales}

At subgalactic scales, the velocity dispersion of stars in the Pal~14 globular cluster is consistent with Newtonian dynamics and no DM, which is in accordance with the CDM paradigm, but might be problematic for MOND\cite{Jordi}: however, more studies of this kind are required since (i) the analysis was based on very few stars, for which one gaussian has been fitted to the total velocity distribution (while a Newtonian core and MONDian outskirts might be expected in MOND), (ii) no anisotropy was considered (which could severely affect the conclusions depending on the position of the tracer stars in the cluster), (iii) the possible effect of mass segregation was ignored, and (iv) the MOND prediction was based on a purely circular orbit for Pal~14 around the Milky Way. Let us note that this last caveat also applies to the long-standing problem of the MOND overprediction of the Roche lobe volume of globular clusters\cite{Tian}.

\subsection{Extragalactic scales}

At extragalactic scales, the CDM paradigm is extremely successful, while MOND still needs large amounts of unseen mass in galaxy clusters, essentially in the central parts, leading to a total discrepancy of a factor 2-3 in the outskirts\cite{AFD}. For large-scale structure (LSS), no simulations comparable to what is done in CDM have been performed, notably because of the absence of a fundamental theory underpinning the MOND paradigm, and because of the unknown nature of the missing mass in galaxy clusters. Finally, present-day Lorentz-covariant extensions of MOND\cite{Bekenstein,Skordis,Li} have not been able to produce a high third-peak in the angular power spectrum of the CMB without producing too much power at large angular scales, or without adding a substantial non-baryonic hot dark matter component\cite{Angus2} (a HDM component which could also solve the galaxy cluster problem of MOND, as well as the problems of ellipticals at the center of clusters). 

\begin{center}
\begin{table*}
\begin{tabular}{ |l |c| c| }
\hline 
 &  {\bf CDM} & {\bf MOND} \\
\hline
HSB & $-$ & $+$ \\
\hline
LSB & $- -$ & $++$ \\
\hline
dIrr & $- -$ & $++$\\
\hline
TDG & $- -$ & $++$ \\
\hline
dSph & $-$ & $-$ \\
\hline
Ellipticals & $-$ & $-$ \\
\hline
Globular Clusters & $+$ & $-$ \\
\hline
Galaxy Clusters & $+$ & $- -$ \\
\hline
LSS & $++$ & $-$ \\
\hline
CMB & $++$ & $- -$ \\
\hline
\end{tabular}
\end{table*}
\end{center}

This overview thus leads us to the following conclusions: while CDM is successful at extragalactic and subgalactic scales, it fails at galactic scales, where it requires an extreme and unrealistic fine-tuning to explain the observations; MOND, on the other hand, is an extremely powerful and  predictive phenomenological law in rotationally supported systems, but it fails in {\it some} pressure-supported systems.

\section{Baryon-DM interactions to reproduce the successes of both MOND and CDM?}

In view of the detailed comparison of the successes of MOND and CDM in the above section, we are naturally led to the idea that both of them could be only effective descriptions valid in different regimes. The first step in such a direction could be a theory which would reduce to General Relativity (GR) plus cold and weakly interacting Dark Matter at extragalactic scales, but that would produce a MONDian behavior for baryons through non-trivial dynamics of the dark sector at galactic scales. However, in view of the above comparison, it should be clear that such a theory would only be a first rough characterization of the desired phenomenology. It is nevertheless already sufficient in order to discuss some non trivial and model-independent facts. We indeed expect the generalized Dark Matter picture to involve Milgrom's acceleration $a_0$ as a parameter, be it a fundamental or effective constant, and also a mass scale, of the order of the TeV, to describe CDM in the appropriate regime. These two basic parameters of the theory may however not be enough to disentangle between galaxy clusters and (spiral) galaxies. As a matter of fact, the internal accelerations within rich clusters and within spirals are almost the same, so that without any other order parameter in the theory, we will have to consider that the same physics applies for both these types of objects, in contradiction with the desired phenomenology. 

The point is that, in the unified picture we are looking for, there is no reason why $a_0$ should play both the role of a characteristic acceleration at which dynamics departs from Newton, and the role of an order parameter that separates the MONDian regime from the CDM one. In the case where $a_0$ plays this double role, i.e. when the unified picture only involves $a_0$ and the mass $m$ as parameters, the theory generically implements a transition from MOND to Newtonian (or GR) dynamics plus Dark Matter, instead of the usual transition from MOND to Newtonian dynamics without Dark Matter. For instance, it has been suggested that DM could carry a space-like four-vector in analogy with dipoles\cite{Blanchet}: in this formalism, the Lagrangian in the matter action for DM has an ordinary mass term for a pressureless perfect fluid, but also has terms depending on a vector field (representing the dipole four-vector moment carried by the DM fluid) and on its covariant time derivative. This implies that the contribution of the DM fluid to the energy density involves a monopolar and a dipolar term: the monopolar term can then play the role of CDM in the early Universe, and the dipolar term takes over in galaxies and creates the MOND phenomenology through what can be called ``gravitational polarization". The success of this model relies on a ``weak-clustering" hypothesis, namely that, in galaxies, the DM fluid does not cluster much and is at rest because the internal force of the fluid precisely balances the gravitational force, in such a way that the polarization field is precisely aligned with the gravitational one in order to create the MOND effect. 

More generally, in the absence of a weak clustering hypothesis, these kinds of models where $a_0$ plays a double-role generically predict the presence of Dark Matter (Cold if the mass $m$ is large) within every collapsed region where internal accelerations are high with respect to $a_0$. This is the case for central regions of both HSB and rich galaxy clusters. As it is however well-known, there is no need for (and thus not much room for) Dark Matter in the central regions of HSB (until radii of order several kpc). Hence such a mini-halo of Dark Matter predicted by this kind of theory should always be, for some reason, of negligible mass (weak clustering hypothesis). On the contrary, we have seen that a large amount of Dark Matter is still needed in MOND at cluster scales. This missing mass could then perfectly well originate from the corresponding mini-halo in the central parts of galaxy clusters (where internal accelerations are several times $a_0$). It is however generally unclear why the mini-halo should have a negligible mass within galaxies but a large one within clusters. Whether this is indeed the case or not can only be checked through simulations of structure formation in the unified $\Lambda$CDM - MOND models.

There are then two possible ways out. One is to consider that the mass of the Dark Matter field is actually small, hence being a Hot Dark Matter (HDM) candidate. It is then natural that the HDM halo at the center of HSB weighs negligibly, whereas it can cluster significantly at larger scales. Cosmology of course becomes GR+HDM, which is known to have some problems, although some of them may be avoided by MONDian effects at late times (enhanced gravity during the formation of structures). 

The other way around rests on the initial program, i.e. designing a framework in which clusters must be CDM dominated (no MONDian effects at all), whereas spirals are MONDian. This way, the mass discrepancy within clusters is explained as in the CDM paradigm, and the weak clustering property is only needed for HSB. Such theories, as stressed above, require the introduction of new scales in order to separate clusters from the central parts of HSB (as the internal acceleration cannot play this role). The size of these collapsed systems may be a relevant quantity, as they differ by orders of magnitude. However, apart from using a mass term (already discussed hereabove), the size of collapsed objects is a non-local quantity, which is difficult to obtain as a combination of local variables (fields) within a local theory. Their total mass may also be a relevant parameter. This is already more natural than size, as there is at least one example of a theory (which, however, is unstable and thus ill-defined) in which MONDian or Newtonian behavior do not only arise at some critical acceleration, but also for some critical mass (Phase Coupling Gravitation\cite{Sanders88}). Another possibility, in view of our comparison of the respective merits of MOND and CDM in various systems in Section~2, would be to simply distinguish MONDian rotationally supported systems from CDM pressure-supported systems, but this distinction is also impossible to express in a local field-theory. So, in the end, the simplest and most natural candidate to separate MONDian systems from CDM ones seems to be the local baryonic medium density (or gas density): the typical medium density in galactic disks is $\rho \sim 5\times 10^{-22} {\rm kg/m}^{3}$, whereas the typical density in galaxy clusters is $\rho \sim 10^{-25} {\rm kg/m}^{3}$. Hence we may introduce a new parameter of density, $\rho_0$, of order $10^{-24} {\rm kg/m}^{3}$, i.e. of the order the MONDian density $a_0^2/G c^2$ up to a pure number, in order to separate galaxy clusters from spiral galaxies. A theory with such a new order parameter would reduce to GR+CDM everywhere except in low acceleration {\it and} high medium density environments (low density in the sense of the trace of the stress-energy tensor for baryons $T_{\rm mat}$). With such a new order parameter, HSB, LSB and TDG would be MONDian, but fluffy globular clusters in the outskirts of galaxies may be either MONDian or Newtonian depending on their gas density, while gas poor ellipticals and dwarf spheroidal satellites of the Milky Way would also behave according to CDM rather than to MOND. Let us also note that, owing to the hierarchical scenario, clusters collapse after galaxies, and therefore in a very low density environment. This ensures that they are indeed CDM dominated through their whole history.

We defined\cite{nous} a whole class of models that are able to implement the basic ideas we have discussed so far. The models all consist of standard GR including a cosmological constant, minimally coupled standard matter fields $\psi$, a simple massive scalar field $\chi$ acting as Dark Matter, and a new term in the action describing the interaction between standard and dark matter fields. Hence the action reads
\begin{equation}
\label{Action1}
S = S_{\textrm{EH}}[g_{\mu\nu}] + S_{\textrm{Mat}}[\psi, g_{\mu\nu}]+S_{\textrm{DM}}[\chi, g_{\mu\nu}]+S_{\textrm{Int}}[\chi,\psi,g_{\mu\nu}].
\end{equation}
The interaction term is defined as 
\begin{equation}
\label{SInteraction}
S_{\textrm{Int}}[\chi,\psi,g_{\mu\nu}]= \frac{1}{2 c} \int \sqrt{-g} d^4x \, T_{\textrm{Mat}}^{\mu\nu} h_{\mu\nu},
\end{equation}
where $h_{\mu\nu}(\chi, \psi, g)$ is some rank-two symmetric tensor that depends on $\chi$ and its derivatives\cite{nous}. It is then clear that the model reduces exactly to $\Lambda$CDM in vacuum (for baryons), and approximately when the interaction is negligible. The tensor $h$ can moreover be defined in such a way that, close to any distribution of baryonic matter the Dark Matter fields couples to standard matter and gives it a MONDian dynamics, whenever its acceleration is small with respect to $a_0$. These models thus describe the phenomenology we were discussing above, where no additional order parameters are present besides $a_0$ and $m$. Varying $m$ then describe either a ``MONDian HDM" or a ``MONDian CDM". Introducing the density as a new parameter is straightforward. It suffices to multiply the above tensor $h$ by some function of the density, $F(\vert T_{\textrm{Mat}} \vert/\rho_0)$, with the relevant asymptotic behavior: $F$ should vanish for small density but saturate to $1$ at large density.

This class of simple models should be viewed only as a very first step, as they suffer from naturalness problems, and are still far from explaining the physical origin of $a_0$ or any possible link with dark energy. Many things should still be investigated within such a class of models: (i) the high medium density in the disks of spiral galaxies should produce a MONDian metric, but in the halo, nothing a priori prevents the dynamics to behave according to CDM: the solution to this problem should lie within the boundary conditions in a static situation, and within numerical simulations of galaxy formation; (ii) in the high-acceleration regime close to the center of HSB, nothing a priori prevents the formation of a mini-halo of dark matter discussed above: solutions to this problem may rest on numerical simulations of galaxy formation or by making the mass of DM also vary as a function of the medium density; (iii) when deriving the equations of motions for a simple toy-model in which the baryons are represented by a standard scalar field, one notices some mixing between the baryon and DM fields: it will be of high interest to investigate the consequences of such mixings on a possible substantial creation of baryonic matter from DM (and conversely) when the interaction term becomes non-negligible in the process of galaxy formation. 

Finally, let us note that the fact that the medium density (or any other parameter) is indeed a relevant parameter controlling the transition from MOND to CDM in low acceleration regimes is in principle a testable hypothesis: falsifying it would require to find a low acceleration and high medium density stellar system not strictly obeying Milgrom's law.

\end{document}